\documentclass[10pt, journal, compsoc]{IEEEtran}

\usepackage[utf8]{inputenc}
\usepackage[T1]{fontenc}

\usepackage{amsmath}
\usepackage{amssymb}
\usepackage{amsfonts}

\usepackage[noend, algoruled, linesnumbered]{algorithm2e}

\usepackage[pdftex]{graphicx}
\usepackage{svg}
\usepackage{subfig}
\usepackage{paralist}
\usepackage{lineno}
\usepackage[nocompress]{cite}

\begin{document}

\title{Using Surrogate Models and Data Assimilation for Efficient Mobile Simulations}

\author{%
  Christoph Dibak,
  Wolfgang Nowak,
  Frank Dürr,
  Kurt Rothermel
  \IEEEcompsocitemizethanks{
    \IEEEcompsocthanksitem C. Dibak, F. Dürr, and K. Rothermel are with the Institute of Parallel and Distributed Systems (IPVS), University of Stuttgart, Germany, Email: \{dibak, duerr, rothermel\}@ipvs.uni-stuttgart.de
    \protect\\
    \IEEEcompsocthanksitem W. Nowak is with the Institute of Modelling Hydraulic and Environmental Systems (IWS), University of Stuttgart, Germany,
    Email: wolfgang.nowak@iws.uni-stuttgart.de
  }
}

\markboth%
  {Preprint}%
  {Dibak \MakeLowercase{\textit{et al.}}: Using Surrogate Models and Data Assimilation for Efficient Mobile Simulations}

\IEEEtitleabstractindextext{%
  \begin{abstract}
    Numerical simulations on mobile devices are an important tool for engineers and decision makers in the field.  However, providing simulation results on mobile devices is challenging due to the complexity of the simulation, requiring remote server resources and distributed mobile computation.  The additional large size of multi-dimensional simulation results leads to the insufficient performance of existing approaches, especially when the bandwidth of wireless communication is scarce.  In this article, we present an optimized novel approach utilizing surrogate models and data assimilation techniques to reduce the communication overhead.  Evaluations show that our approach is up to $6.5$ times faster than streaming results from the server while still meeting required quality constraints.
  \end{abstract}

}

\maketitle

\IEEEdisplaynontitleabstractindextext

\IEEEraisesectionheading{\section{Introduction}\label{sec:introduction}}

\IEEEPARstart{A}{s} of today, numerical simulations are heavily used in engineering and decision making.  While today's simulations are mostly executed on powerful stationary computers, many use-cases require simulation results to be available in the field including on-site information or interaction with the simulation.  Enabler for such mobile simulations are modern powerful mobile devices, emerging augmented reality headsets for interaction, and more and more sensors being integrated into the Internet of Things.

Engineers facing unplanned or unforeseeable situations would benefit from such mobile simulations allowing for faster decision making.  As an example, consider an engineer in the field who has to find a solution for placing a hot exhaust tube during deployment of a machine in a factory.  Using her augmented reality headset and mobile simulations, the engineer directly sees the heat of the surface of the tube and its surrounding materials as if the machine would be operational.  To cover different situations, the engineer can change parameters of the simulation, e.g., the airflow surrounding the tube.  The application enables the engineer to see the heat distribution even in complex geometrical regions, e.g., bends, and allows her to place the tube according to surrounding material.  Additionally, parameters from sensors can be integrated into the simulation, e.g., to include data from similar machineries deployed elsewhere.

The main challenge for implementing such mobile simulations is the complexity of the simulation combined with resource-poor mobile devices.  Mobile devices are about $10$ times slower than servers for solving numerical problems~\cite{Dibak2015}.  Even worse, for very high-quality solutions, the battery-powered mobile device might run out of memory or energy and is therefore unable to provide any solution.  Therefore, powerful server resources need to be involved to enable high-quality simulations on mobile devices.  However, simply streaming simulation results from remote servers to mobile devices has been found to be inefficient due to the large size of the results and the latency of the communication~\cite{Dibak2017a}.  Efficient approaches therefore utilize both, communication and computing resources of the mobile devices and carefully consider quality degradation to reduce latency of the computation.

Recent literature suggests using code-offloading techniques for dynamically distributing a mobile application between remote server and mobile device~\cite{Cuervo2010, Ra2011, Chun2011, Gordon2012}.  Code-offloading typically splits the application into modules that will be dynamically placed on either the mobile device or the server.  For this placement, data dependencies between modules have to be taken into account possibly resulting in communication overhead, e.g., one module on the mobile device requires data from another module on the server.  However, numerical simulations require lots of state information during the execution, making distributed execution between mobile device and server constantly unpleasant.  Therefore, code-offloading would also result in the two inefficient options discussed above:  either compute everything on the mobile device or stream everything from the server.  Additionally, code offloading is application agnostic and therefore misses the opportunity to exploit specific properties of the simulation to gradually reduce quality while still providing results within user-defined quality bounds.

Our recently proposed methods for mobile simulations are based on existing numerical methods, namely model order reduction, to provide efficient distributed execution between a server and a mobile device~\cite{Dibak2017a, Dibak2017b, Dibak2018a}.  Model order reduction generates a reduced model in a pre-computation step.  The reduced model is built to provide fast approximate solutions and can be executed on the mobile device.  While the generation process for the reduced model is compute-intensive, it can be executed on a remote server, which can also adapt the reduced model when necessary.  However, these approaches only consider time-independent problems, only work for similar behavior of the simulation model that can be expressed using only a few parameters, and are less flexible due to the pre-computation step.  Consequently, time-dependent simulations with a high number of parameters require new methods for distributed mobile execution.

In this article, we propose a novel framework for the distributed execution of generic time-based simulations between mobile device and server based on, so-called, surrogate models.  Surrogate models are computationally simplified models of a simulation model providing approximate solutions for the reference model.  For instance, the surrogate model can use another discretization, i.e., the simulated system has lower resolution in time or space, it can be based on different physical properties, e.g., neglecting physical properties having only little effect on the result, or even use time-dependent model order reduction techniques.  Intuitively, our approaches will execute the surrogate model on the mobile device and both models, the reference model and the surrogate model, on the server.  Having the results of both models available, the server can decide if the quality available from the surrogate model on the mobile device is sufficient without any communication overhead.  This way, the server can send updates to the mobile device only when necessary to ensure quality constraints for the user.  The remaining challenge is then to integrate updates into the surrogate simulation model.  To solve this problem, one of our approaches will use data assimilation techniques, namely the ensemble Kalman filter, which allows to drastically reduce the size of the updates.

In detail, our contributions in this article are as follows:
\begin{inparaenum}[(1)]
  \item Analysis of the problem for providing time-dependent simulation data to mobile devices;
  \item an approach based on surrogate models where only required parts of the simulation are streamed from a remote server;
  \item an approach based on surrogate models and data assimilation that supports partial updates and therefore significantly reduces the need for high data rates;
  \item evaluation for different scenarios based on real measurements in cellular networks and on a popular system-on-chip platform, as typically used in mobile applications.
\end{inparaenum}

The rest of the current article is structured as follows: Section~\ref{sec:related-work} briefly discusses related work before Section~\ref{sec:system-model} introduces the system model including the mobile environment and the simulation model.  Section~\ref{sec:problem-statement} introduces the problem statement before we introduce the basic streaming approaches in Section~\ref{sec:stream-approach}. Section~\ref{sec:full-update-approach} introduces the full update approach utilizing computation and communication, followed by Section~\ref{sec:partial-update-approach}, which introduces the partial update approach using data assimilation techniques to reduce required bandwidth.  In Section~\ref{sec:evaluation}, we discuss the evaluation of all approaches.Finally, Section~\ref{sec:conclusion-future-work} concludes the paper with future work.

\section{Related Work}
\label{sec:related-work}

Before describing our approaches, we first briefly discuss the limitations of existing approaches for providing time-dependent numerical simulations on mobile devices.  Existing approaches can be categorized into code offloading approaches, mobile approximate computing approaches, and our existing approaches for mobile simulations.

In mobile computing, code offloading partitions a mobile application to run parts of the application on remote server resources.  Existing approaches focus on reducing energy consumption~\cite{Cuervo2010} or the latency of the computation~\cite{Ra2011, Chun2011, Gordon2012}.  The general idea is to partition modules of the mobile application depending on the characteristics and dependencies between modules.  Partitioning will result in two sets of modules, where one set will be executed on the mobile device and the other set will be executed on the server.  Data dependencies between models that do not run on the same node have to be communicated between the server and the mobile device.  For increased robustness against the communication link, modules can also be executed on both nodes~\cite{Berg2015}.  However, the initial partitioning depends on profiling of required resources and data dependencies between modules.

While many mobile applications can benefit from code offloading, mobile simulations require different solutions since
\begin{inparaenum}[(1)]
  \item code offloading is application agnostic and does not consider quality of the result;
  \item numerical simulations are hard to modularize, since computation of one state requires large sizes of shared memory between parallel processes; and
  \item states of the numerical simulation have strong data dependencies, hindering partitioning.
\end{inparaenum}
Therefore, code offloading would result in heavily unbalanced executions, where either a single node computes the full simulation, or the network is heavily used to communicate lots of data between modules.  In contrast, our approaches will consider different quality levels of the mobile and the server computation and only communicate when necessary, reducing both, energy on the mobile device and latency of the application.

A framework for quality aware execution between a mobile device and a remote server has been proposed by Pandey et al.~\cite{Pandey2016, Pandey2017}.  Their approach uses a workflow-based representation of computation tasks that yield approximate results.  Computation tasks are profiled offline prior runtime.  During runtime, the previously profiled tasks are offloaded between mobile device and server.  However, while there are quality aware algorithms that can benefit from their framework, it is unsuitable for numerical simulations, since
\begin{inparaenum}[(1)]
  \item the separation of offline and online phase make this approach infeasible for applications that depend on parameters, such as numerical simulations;
  \item the workflow of numerical simulations consists of varying number of tasks for varying quality, which is not considered in their approach; and
  \item as for code offloading, the workflow of numerical simulations consists of a single path of tasks with strong data dependencies, which results in unbalanced offloading.
\end{inparaenum}
In contrast, our approach does not require any offline phase or profiling before runtime and will execute different quality versions of the single-path workflow in parallel on the server and the mobile device to deal with strong data dependencies.

In our previous works, we used the reduced basis method (RBM) that reduces the complexity of simulations to provide faster results~\cite{Dibak2017a, Dibak2017b, Dibak2018a}.  The general idea is to pre-compute a simpler model that provides only approximate results at much lower cost.  Using the RBM we were able to significantly reduce the runtime and energy consumption.  However, our existing approaches only focused on stationary simulations which are time-independent and therefore are unusable for time-dependent simulations.  Additionally, approaches presented in this article will perform much better for higher number and broader range of parameters.

Other approaches by the authors focused on increasing the robustness of streaming results from a remote server to a mobile device under disconnections~\cite{Dibak2015}.  Using a statistical model, we were able to predict the duration of disconnections and therefore decide on the computation on the mobile device.  Such approaches are orthogonal to the work presented in this article and can be used to make approaches more resilient against disconnections.

\section{System Model}
\label{sec:system-model}
\graphicspath{{figs/model/}}

This section introduces our system model for dynamic offloading of time-dependent numerical simulations.  We first describe our model for the time-dependent simulation and then provide our model of the mobile environment, consisting of mobile device, remote server, and wireless communication network.

\subsection{Time-Dependent Simulations}

\begin{figure*}[bt]
  \centering
  \includegraphics{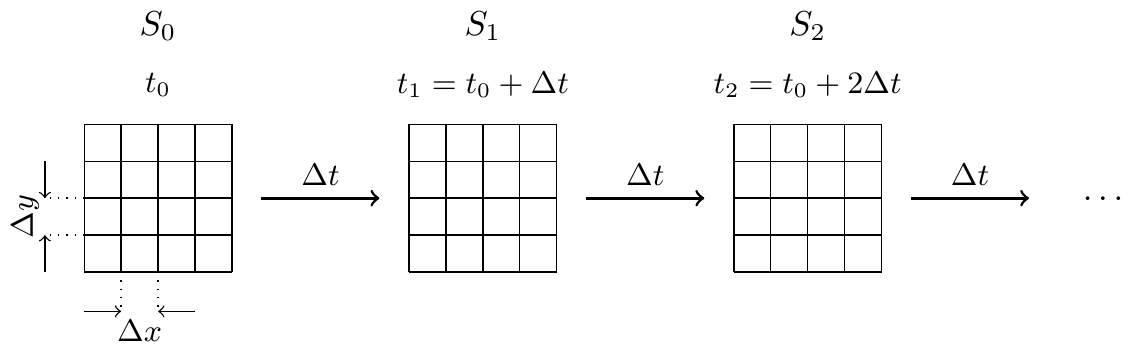}
  \caption{Time discretization and space discretization of numerical simulations.}
  \label{fig:discretization}
\end{figure*}

Time-dependent numerical simulations are based on differential equations describing the behavior of the system w.r.t.\ continuous time and space.  Such equations need to be discretized in order to be solved.  Time-discretization divides the continuous time into $n_t + 1$ time steps.  Each step represents the system state $S_i$ at fixed time $t_i$.  For simplicity, we assume that the time of the first step is $t_0 = 0$ and the time for the last step is $t_{n_t} = 1$.  Then, the time resolution is $\Delta t = 1 / n_t$ (cf. Fig.~\ref{fig:discretization}).

Next, we first describe how the simulated system at each time step is discretized in space, how the transition between time steps is implemented, and how the computation can be optimised to provide computationally cheaper approximations of the simulation problem.

\subsubsection{Representation of Time-States and Transition Between States}

Time-states $S_i$ represent the state of the simulated system at discrete points in time.  While space is defined continuously in the differential equation, it also needs to be discretized.  To this end, the system is only observed at fixed points in space, e.g., at points forming a grid with mesh width $\Delta x$ (cf. Fig.~\ref{fig:discretization}).  Values of the simulation at these points form a vector.  The size of the vector depends on the spatial discretization.  If finer discretization is required, the size of the vector is increased.  The size of the vector later also depends on the complexity of the computation.

Transition between time states is implemented by solving an algebraic problem in a numerical solver.  The output of the solver is the state vector of the next time state $S_{i+1}$.  Input into the solver is the old state vector $S_i$ and problem specific information, e.g., a problem matrix and a vector forming the algebraic problem.  Typically, there is a choice between multiple classes of algebraic problems for the same differential equation leading to different trade-offs between quality and complexity of the computation.  For instance, simulating heat propagation using the heat equation yields various discretization methods that can be generalized into two classes:  implicit methods and explicit methods.  While implicit methods are computationally more expensive they provide better quality than explicit methods.  Such decisions on the trade-off between quality and complexity motivate the use of surrogate models.

\subsubsection{Reference Model and Surrogate Model}

\label{sec:reference-model-and-surrogate-model}

We assume two different implementations of the model, a reference model and a surrogate model.  The reference model defines the ``ground truth'' of the simulation.  It is defined over fine-grained discretization grids enabling accurate predictions of future system states.  While it provides accurate results, it is expensive to compute.  On the other hand, the surrogate model is computationally less expensive while providing only lower quality than the reference model.  Surrogate models can be obtained by using an explicit method rather than an implicit method or by changing $\Delta x$ of the discretization grid to have a lower space resolution.

We will later compare the reference model and the surrogate model at the same time step.  To compare results of both models, the vector of the reference model has to be mapped to the same dimensionality as the vector of the surrogate model.  We assume that this mapping is provided by a transformation matrix $T_{R\to S}$.  Additionally, to simplify the notation for comparison between models, we assume that time-discretization of the reference model and the surrogate model is the same.  However, the reference model could also be configured to compute multiple, say $n_{\textit{ref}}$ steps for one surrogate step.  This way, the reference model will have a time discretization with $\Delta t_{\textit{ref}} = \Delta t / n_{\textit{ref}}$ and we are able to compare results of the reference model and the surrogate model every $n_{\textit{ref}}$ time steps.

\subsubsection{Mixing Simulation Models for Approximate Solutions}

The solutions of one simulation model form a chain of time steps (cf.~Fig.~\ref{fig:discretization}).  To provide better quality, the chains of the reference model can be used to update the surrogate model chain.  Each of these updates forms a new chain of approximate solutions.  For instance, the surrogate model is updated at time step, say 5, to set its state to the state of the reference model.  The resulting chain of simulation results may be significantly different compared to the original surrogate simulation chain without the update.

\subsection{System Components}

\begin{figure}[b]
  \centering
  \includegraphics{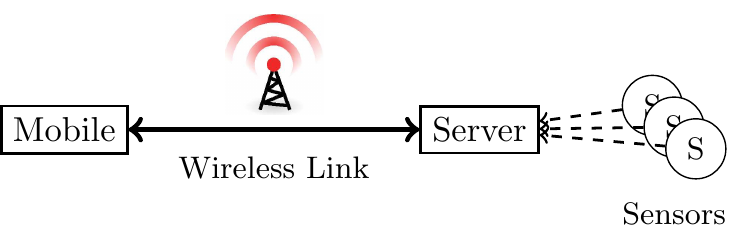}
  \caption{The two compute nodes, server and mobile device, and connected sensors.}
  \label{fig:nodes-sensors}
\end{figure}

To compute results of the numerical simulation, the system consists of two compute nodes, namely the mobile device and the server.  The server is located in a central location in the network and receives data from sensors (cf. Fig.~\ref{fig:nodes-sensors}).

The mobile device is carried by the user.  The user directly interacts with the mobile device and requests simulation results.  The mobile device has an energy-efficient but slow processor.  In contrast to the server, it is very resource-limited and depends on batteries providing limited energy.

The server receives data from cloud-connected sensors and collects and stores relevant data to form the initial state for the simulation.  Therefore, the initial state for the simulation is only available on the server and needs to be communicated to the mobile device before any simulation model can be executed.

Server and mobile device are connected via wireless communication, e.g., 3/4G cellular networks or IEEE 802.11 WiFi.  Wireless communication is subjected to dynamic latency and throughput, which might be very low in some cases.

\section{Problem Statement}
\label{sec:problem-statement}

After describing the system model, this section describes the problem statement.  We first define quality for approximate solutions and then define the optimization goal of the system.

Quality of approximate solutions is defined by comparing approximate solutions to the reference model using a user-defined norm $\lVert \cdot \rVert_U$.  Let $S_i^A$ denote the approximate solution and $S_i^R$ denote the reference solution for time step $i = 0, \dots n_t$.  The quality of time step $i$ is then defined as $q_i^A = \lVert S_i^A - T_{R\to S} S_i^R \rVert_U$, where $T_{R\to S}$ is the transformation matrix between reference model results and surrogate model results (cf.~Sec.~\ref{sec:reference-model-and-surrogate-model}).  The overall quality of the approximate solution for all time steps is then defined as
\begin{align}
  Q_A
  & = \max_{i = 0, \dots, n_t} q_i \\
  & = \max_{i = 0, \dots, n_t} \left \lVert S_i^A - T_{R\to S} S_i^R \right \rVert_U.
\end{align}
One example for the user-defined metric $\lVert \cdot \rVert_U$ is to compare approximate solution and reference model solution by the maximum difference at any point, e.g., the maximum temperature difference of a heat simulation.

Having defined quality, we can now define the overall goal of the system.  The goal of the system is to minimize the latency until approximate solutions are available on the mobile device.  Approximate solutions have to fulfill quality constraints given by the user, i.e., the user provides $Q_{\max}$ and the solution has to fulfill $Q_A \leq Q_{\max}$.  This way, the user can define the maximum difference of the approximate solution to the reference simulation and the system provides an approximate solution as fast as possible.

\section{Basic Streaming Approaches}
\label{sec:stream-approach}

We briefly describe the straight forward streaming approaches.  These approaches only serve as baseline and for comparison to our approaches that will be presented in the next sections.  Streaming approaches compute all steps of the simulation on the server and communicate results to the mobile device.  We introduce two approaches, the \emph{simple stream approach} and the slightly more sophisticated \emph{advanced stream approach}.

The simple stream approach computes the reference simulation on the server and communicates all steps to the mobile device.  Therefore, all results on the mobile device have the best possible quality and $Q_A = 0$.

The advanced stream approach also computes the reference simulation on the server.  However, it will reduce the quality of the simulation states before they are sent to the mobile device.  In particular, it will reduce the resolution of the simulation to the surrogate model discretization.  This way, the quality is still $Q_A = 0$, while the volume of data communicated over the wireless communication link is significantly reduced.

While the simple stream approach represents the result of an unbalanced partitioning, which could be the result of code offloading for the simulation problem, the advanced stream approach is able to reduce the communication overhead at no quality loss.  However, the advanced stream approach still has to communicate all simulation steps over the network.  The following sections will introduce our approaches, which require much lower communication overhead over the stream approaches.

\section{Full Update Approach}
\label{sec:full-update-approach}
\graphicspath{{figs/full-update/}}

While the previously introduced stream approaches are able to meet any quality requirements, they do not consider the mobile device for computing simulation results.  This section therefore introduces the full update approach, which reduces the requirements on the wireless link as it uses computation on the mobile device.

The general idea of this approach is to execute the same simulation on the surrogate model simultaneously on the mobile device and the server.  Thus, the server knows which (approximate) results have been calculated by the mobile device using the surrogate model.  By comparing to the exact solution of the reference model, the server can send updates to the mobile device at selected states, whenever the surrogate model yields results of insufficient quality.

\begin{figure}[t!]
  \centering
  \includegraphics{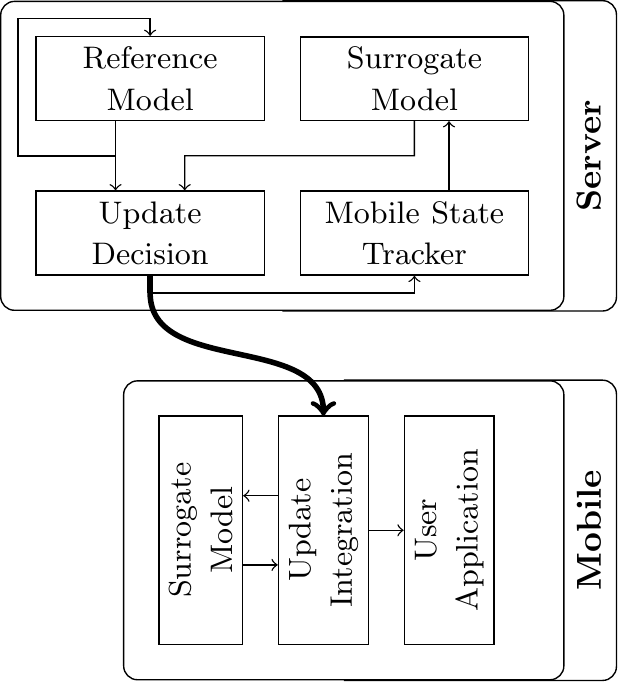}
  \caption{Overview of the full update approach}
  \label{fig:full-update-approach}
\end{figure}

Figure~\ref{fig:full-update-approach} depicts the components for the full update approach.  Reference model and surrogate model are models of the simulation that implement time transition of the states.  The update decision component will decide whether to send an update to the mobile device.  The mobile state tracker holds the last known state of the simulation on the mobile device for the previous state.  On the mobile device, the update integration component combines possible outputs of the surrogate computation.

While the reference model and the surrogate model have already been introduced in the previous sections, we will explain the remaining components in the following subsections.

\subsection{Mobile State Tracker}

The mobile state tracker provides the previous state of the mobile device on the server.  As the surrogate model is deterministic and will return the exact same result as on the mobile device, the server can use the result even before it has been computed on the mobile device.

For the initialization, the initial state needs to be communicated to the mobile device.  The mobile state tracker will then be initialized with the same initial state.

\subsection{Update Decision}

The update decision is based on the requirements of the user.  To this end, the update decision component receives the current state of the reference model and the surrogate model.  It computes the difference of the states after transforming the reference state to the same spatial discretization grid as the surrogate state.  Afterwards, it will check whether the quality of the result of the surrogate model is sufficient.  If it is sufficient, it will send a quality certification message to the mobile device.  If it is not sufficient, it will send an update of the vector representing the current reference state to the mobile device.

Notice that before sending, the update is transformed to the spatial discretization level of the surrogate model since this provides the quality such that the mobile device can continue calculating future states from the updated model.

The update decision component will also update the tracked state on the server.  If a certification message was sent, it will use the result from the surrogate model.  If an update was sent, it will update the mobile state with the result from the reference model.

\subsection{Update Integration}

The update integration component is executed on the mobile device and receives messages from the server.  It may invoke the surrogate model and provides the result as current simulation state to the user application.

If the update integration component receives a certification message, it will use the result from the surrogate model and provide the result to the user application.  If it receives an update message, it parses the state and provides it directly to the user application since the result then is directly derived from the reference model.  To this end, the update integration module always has to wait for the next message from the server.  If no certification message is available, it cannot provide any result to the user application since the result of the surrogate model might not provide sufficient results.

There are optimistic and pessimistic implementations of the full update approach.  The optimistic implementation assumes that no update has to be made and the computation using the surrogate model is sufficient.  The pessimistic implementation assumes that the surrogate model will not provide sufficient quality.  To this end, the optimistic approach will always start the computation of the next state in the background, whereas the pessimistic approach only computes on request by the server.

Compared to the stream approach, the full update approach will reduce the traffic of the wireless communication link as certification messages will be much smaller than streaming results.  The traffic of the network now depends on the accuracy of the surrogate model.  However, this approach adds slightly more overhead on the server side, which now has to compute the surrogate model in parallel to the reference model.  In the next section, we will see how we can further reduce the traffic of the network by just sending partial updates.

\section{Partial Update Approach}
\label{sec:partial-update-approach}
\graphicspath{{figs/partial-update/}}

\begin{figure*}[bt]
  \centering
  \includegraphics{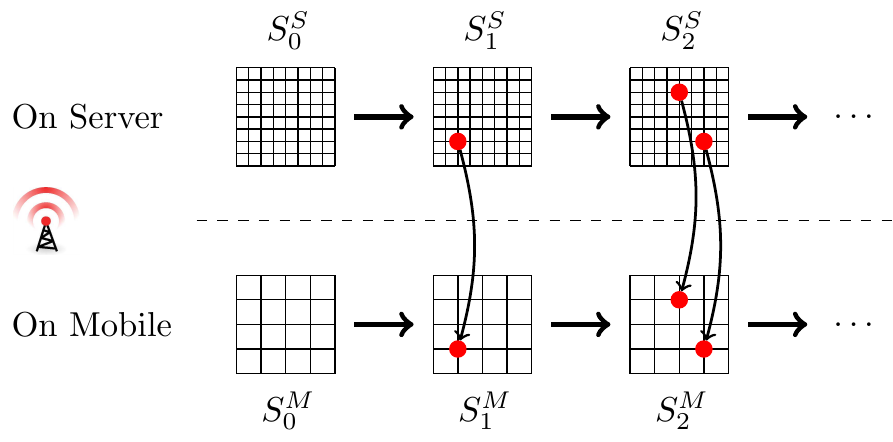}
  \caption{Partial point updates.}
  \label{fig:partial-updates}
\end{figure*}

The previously introduced full update approach always has to communicate full state updates for surrogate states violating quality requirements.  This results in a huge communication overhead even in cases where only a small part of one surrogate state violates quality constraints.  In this section, we therefore introduce our partial update approach, which is based on the full update approach, but will only update parts of the vector representing the approximate solution from the surrogate model (cf. Fig.~\ref{fig:partial-updates}).

Updating single values of the simulation is not straight forward since the simulation model might be sensitive to external changes of the simulation state.  For instance, if we consider a heat simulation, the simulation model and numerical calculation assumes the solution to be continuous, while when randomly updating values, the solution becomes discontinuous, i.e., updated values add sharp edges to the simulation state.  Such discontinuities lead to numerical instabilities which would never occur in normal calculations for the simulation and which the model might not be able to recover.  Therefore, more sophisticated approaches are required.

To reduce the discontinuity when updating single values, we use data assimilation techniques.  Data assimilation emerged from weather simulations, where sensor data updates the simulation state, which leads to similar problems~\cite{Law2015}.  To prevent such problems, data assimilation techniques identifies the correlation between parts of the simulation state.  When updating one value, data assimilation uses this correlation and updates all correlated values respectively.  Therefore, the number of discontinuities is highly reduced and the simulation model quickly recovers from single point updates.

The idea of the partial update approach is to apply data assimilation and treat single point updates as sensor observations.  In this case, sensor observations are perfect, since they are taken from the reference simulation, which is our ground truth.  This simplifies the calculation of data assimilation methods, since they normally assume inaccurate observations.

In the following, we first briefly describe our data assimilation of choice, the ensemble Kalman filter, before we discuss how the partial update approach changes the update decision and update integration from the full update approach.

\subsection{The Ensemble Kalman Filter}

\begin{figure*}[bt]
  \centering
  \includegraphics{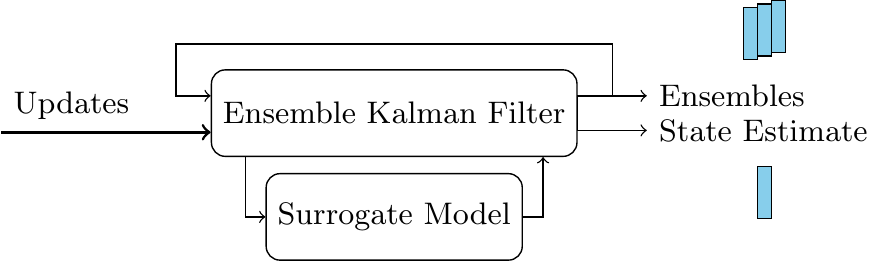}
  \caption{Simplified operating principle of the ensemble Kalman filter.}
  \label{fig:ensemble-kalman-filter}
\end{figure*}

The ensemble Kalman filter (EnKF) provides a solid and frequently applied framework for data assimilation~\cite{Evensen2003}.  The general idea of the EnKF is to use multiple states to track uncertainties (c.f. Fig.~\ref{fig:ensemble-kalman-filter}).  These states are called ensemble members.  Initially, ensemble members are generated using random perturbation of the initial state.  For every simulation state, the next state of the ensemble members is computed using the surrogate model.  The number of ensemble members $n_e$ can be small.  It has been shown that even some complex applications do not require more than $50$ ensemble members~\cite{Houtekamer1998, Keppenne2000}.

\subsubsection{Generation of Ensemble Members}

We generate ensemble members by perturbation of a reference state $S_i$.  To this end, we add a random vector to the initial state to form an ensemble member $e_{i}^{(j)} = S_i + r^{(j)}$.  The random vector $r^{(j)}$ is sampled such that the mean of ensemble members track the state of the reference simulation, e.g., using the standard error between reference and surrogate model.

In order to have the same result available on the mobile device as on the server, the computation has to be deterministic.  To provide random numbers to the EnKF, we therefore use a deterministic random number generator with well-defined seed for the random vector.  As the seed should change for every state, the server chooses a basic seed during initialization.  We then use a deterministic function of the basic seed and the state number to calculate the seed for state perturbation of the current state.

\subsubsection{Combining Simulation Model and Observations}

Combining the state of the surrogate model and partial updates consists of two steps, namely the forecast step and the analysis step.  In the forecast step, the surrogate model is applied for all current ensemble members $E_i = (e_i^{(1)}; \dots; e_i^{(n_e)})$.  This generates the forecast ensembles for the next ensemble members $F_{i+1} = (f_{i+1}^{(1)}; \dots ; f_{i+1}^{(n_e)})$.

Partial updates are communicated as set of pairs $(\textit{position}, \textit{value})$ where, for every updated value, the position of this value in the surrogate state vector is given.  This representation is translated into an update vector $u_{i+1}$ containing just the values and a measurement operator $H_{i+1}$ mapping respective entries of the surrogate state vector to the update vector.

For the analysis step, the next state has to be combined with partial updates $u_{i+1}$ by using the so called Kalman gain $K_{i+1}$.  The Kalman gain defines the sensitivity of the difference of partial update $u_{i+1}$ and forecast state $F_{i+1}$.

The analyzed ensemble members are then calculated as $e_{i+1}^{(j)} = f_{i+1}^{(j)} + K_{i+1} (f_{i+1}^{(j)} - H_{i+1} u_{i+1})$.  The analyzed simulation state as output for the user is the ensemble mean of all analyzed ensemble members.  Further details about the EnKF and the computation of the Kalman gain $K_{i+1}$ can be found in the appendix.

\subsection{Update Decision}

For identification of parts that need updating, we introduce the concept of violation points.  Violation points are points in the result of the surrogate model that violate the quality constraint.  How these points can be calculated depends on the norm used for specifying the quality.  In general, we distinguish between maximum norm and any other norm.

If the maximum norm is used for the quality constraint, every point has a maximum distance to its corresponding point in the reference state.  Violation points are therefore all points that differ too much from the current reference state.

For other norms, e.g., the Euclidean norm, the computation of violation points is slightly more complex and requires an iterative process.  To this end, we build the set of points that require updating.  Initially, this set is empty.  If quality requirements using the current updates cannot be met, the point with the maximum error to the reference model is included in the updates.  This is repeated until the quality requirements are met.

Once the decision on the set of points to update is made, the update is sent over the network.  Additionally, on the server side, the update is applied by the mobile state tracker in order to derive the same state as on the mobile device.  In contrast to the full update approach, the mobile state tracker not only keeps track of the current simulation state on the mobile device, but also of ensemble members expressing the uncertainty of the current state.

\subsection{Update Integration}

The update integration component on the mobile device receives the update from the server.  It holds the current ensemble members of the surrogate model.  To this end, it will calculate the prediction model and prepares all steps in the calculation of the EnKF to provide the next state of the simulation for the user application.

Notice that before using the EnKF, we used the Kalman filter as data assimilation technique.  The Kalman filter tracks uncertainty of the states using a covariance matrix.  This matrix is quadratic to the problem size and therefore much more computationally expensive.

\section{Evaluation}
\label{sec:evaluation}
\graphicspath{{figs/evaluation/}}

The previous sections introduced the full update approach and the partial update approach.  This section evaluates both approaches against the streaming approach described in Sec.~\ref{sec:stream-approach}, which is the state-of-the art for providing simulation results to mobile devices.  In this evaluation, we consider different mobile network setups and different assumptions on the accuracy of the surrogate model.  As benchmark simulation problem, we are using a 2d heat simulation based on the well-known heat equation.  Before describing details of evaluation results, we first introduce the evaluation setup.

\subsection{Setup}
\label{ssec:setup}

We evaluated our approaches on a distributed test bed consisting of a Raspberry Pi 3 as mobile device and a powerful server.  The Raspberry Pi 3 uses a system-on-chip (SoC) hardware similar to the SoCs used by mobile devices.  It features a quad-core Broadcom ARM CPU at 1.2~GHz and 1~GB RAM.  The server is a commodity off-the-shelf server featuring a quad-core Intel Xeon E3 CPU at 3.4~GHz and 16~GB RAM.

We emulated the cellular network connecting mobile device and server using the Linux Kernel Packet Scheduler on both nodes.  To this end, we added queueing disciplines that restrict the data rate using a token bucket filter (TBF) and delaying packets using the netem module.  To set parameters of the TBF and for the delay, we measured the performance of real cellular networks using HSDPA and LTE.  We found that in extreme conditions, data rates can be as low as $50$ kbit/s with around $1$ second latency over longer periods.  However, as we assume data rates to increase in the future and as our approaches are much better for lower data rates, we assume a data rate of $1$ Mbit/s.

Our approaches and the simulation are implemented in Python (version 2.7.13) and NumPy (1.14.3).  To accelerate the computation, NumPy was linked with OpenBLAS (0.2.19), which is available for the server and mobile architecture.  Serialization is implemented using Protobuf (3.5.2), and data was communicated using TCP as transport protocol.  We used background threads and queues in order to send data parallel to processing.  As deterministic random number generator, we use the Mersenne twister sequence~\cite{Matsumoto1998} as implemented in Python.

As simulation problem, we choose the popular and well understood 2d heat equation with Dirichlet boundary conditions.  We implemented two numerical solvers, one using explicit Forward-Time Central-Space (FTCS) discretization and the other using the Alternating Direction Implicit method (ADI) with the Crank-Nicolson method for 1d discretization.  Throughout the evaluation, we used the explicit FTCS implementation as surrogate model and the implicit ADI implementation as reference model.  For the initial state, we choose random values in the interval $[0,1]$ and set the boundary to $0$.

The execution of the simulation depends on many parameters for discretization and accuracy of the numerical model.  We ran our evaluations with different parameters and received results similar to the results reported in this section.  For the final evaluations we used the following default parameters.  We assume the temporal discretization as $\Delta t = 0.0001$ with $100$ states.  The maximum error allowed between reference and surrogate model was $2^{-7}$, i.e., we allowed less than 1\,\% of error between reference model and surrogate model.

To compare different surrogate models, we defined quality levels as uniform grids for space discretization.  This way, we can use different discretization grids and define surrogate models on each of the levels.  To this end, our uniform grid implementation consists of different levels, where each higher level includes points of all lower levels plus points in between of all existing points.  The number of points on this levels quickly grows, e.g., the later often referred level $5$ contains $1089$ points, while level $6$ contains $4225$ points.

As the number of updates required might dependent on the required quality, we will in the following first evaluate the impact of quality onto the number of full state updates and sizes of partial state updates.  We will then evaluate the latency for varying data rate of the network, full update probability, sizes of partial updates, and surrogate problem size.

\subsection{Accuracy of the Surrogate Model}
\label{ssec:accuracy-of-surrogate-model}

The accuracy of the surrogate model impacts two quantities:
\begin{inparaenum}[(1)]
  \item the number of points requiring partial updates and
  \item the distribution of the number of points in states violating quality constraints.
\end{inparaenum}
To this end, we introduced the terms \emph{violation states} and \emph{violation points}.  Violation states are states of the simulation that do not fulfill quality requirements and therefore need updating in the full update approach.  Similarly, violation points are points in one simulation state that are required in partial updates.  Intuitively, violation states and violation points depend on the maximum error that is allowed for the application.  To provide an overview of the distribution of violation states and violation points, we recorded them for different error bounds (maximum error) in the 2d heat equation with random initial state.

\begin{figure}[htb]
  \centering
  \includegraphics[width=\linewidth]{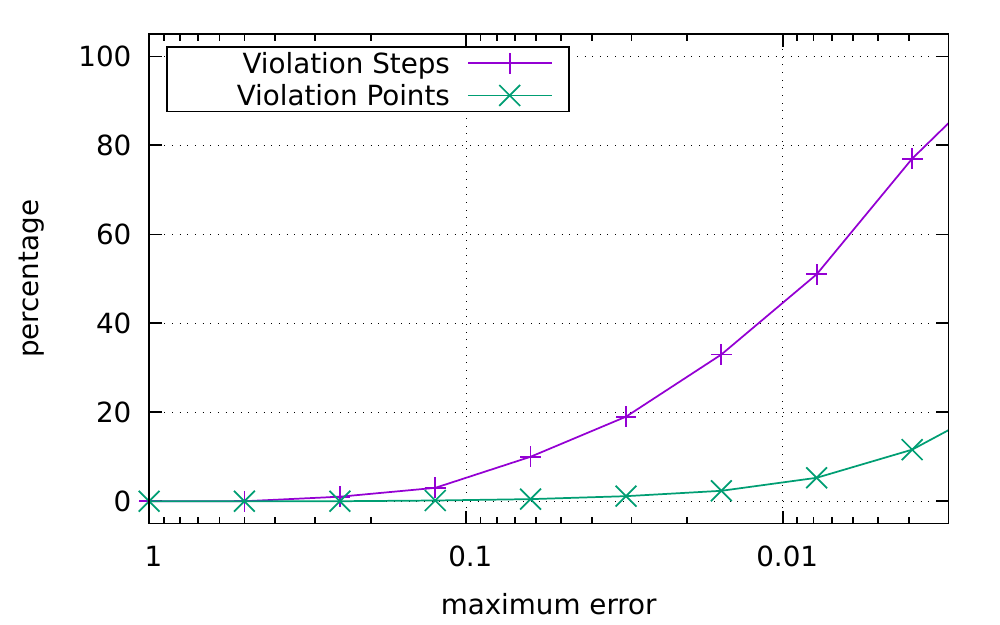}
  \caption{Ratio of violations for full updates and partial updates.}
  \label{fig:violations}
\end{figure}

Figure~\ref{fig:violations} depicts the percentage of states that are required as updates, and percentage of points per state requiring updating for partial updates. While some partial updates require many point updates, the majority of partial updates only require few points.  For maximum error $2^{-7}$, around $50$\,\% of states require updating, while only $5.2$\,\% of points need to be updated.  We therefore assume a state update probability of $0.5$ in the following, if not stated otherwise.  Notice that this reduced the volume of data to be communicated from server to mobile device by $50$\,\%.

\subsection{Impact of Channel Data Rate}

Mobile devices face varying data rates of the wireless communication channel.  Especially in areas with bad signal strength, e.g., indoors in basements, data rates drop to low rates down to $50$ kbit/s.  This is inline with Shannon-Harley Theorem which would require higher bandwidth in cases of higher signal-to-noise ratio to keep constant data rates.

We evaluated the impact of data rates on the latency of our approaches. After setting different rates, we ran our approaches multiple times and recorded the median latency.  All other parameters are as our default configuration introduced in Section~\ref{ssec:setup}.

\begin{figure}[htb]
  \centering
  \includegraphics[width=\linewidth]{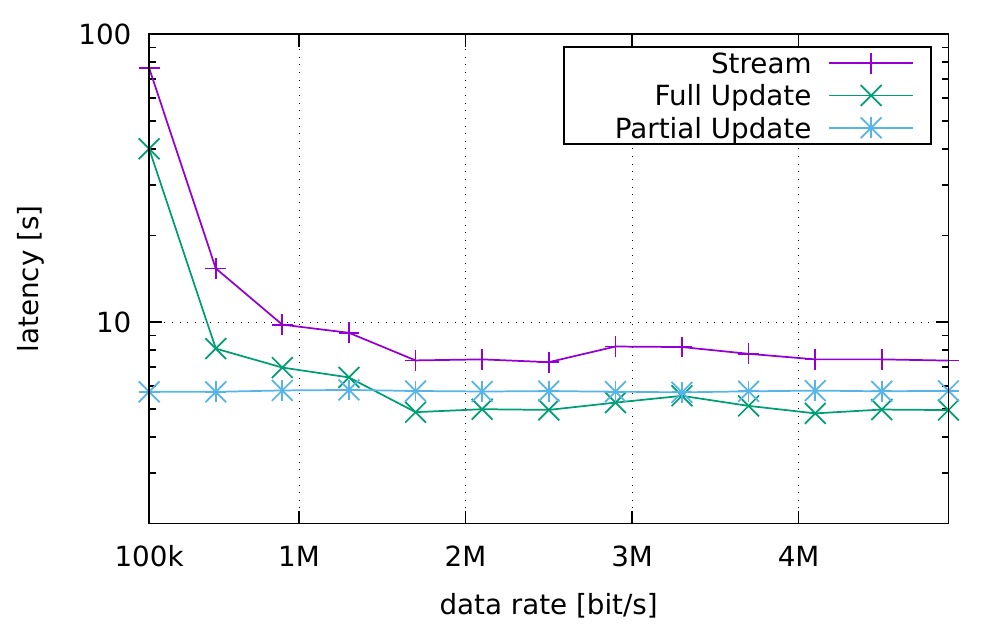}
  \caption{Latency of the three approaches over data rate.}
  \label{fig:throughput}
\end{figure}

Figure~\ref{fig:throughput} depicts results for the three approaches over the data rate of the wireless channel.  As the volume of data to be communicated by our approaches is only a fraction of the streaming approach, both approaches are much faster than streaming.  The full update approach is able to provide results up to $9.4$ times faster than the streaming approach.  The partial update approach, which requires only very little data to be communicated from the server to the mobile device is able to provide results even up to $13.3$ times faster than the streaming approach.  However, for high data rates, the full update approach is marginally faster than the partial update approach since the partial update approach has a higher overhead for data assimilation.  Both approaches have a speedup of $50\,\%$ compared to streaming.  We assume that this speedup is due to the doubled volume of data for streaming and the serialization using state-of-the-art Google Protocol Buffers, which is already considered to be one of the fasted serialization formats~\cite{Sumaray2012}.  In general, data rate has only very little impact on the partial update approach, while it effects stream and full update approach.  For varying data rates, the combined approach is therefore the best choice, while the full update approach might be considered for higher data rates.

\subsection{Impact of Update Probability}

To evaluate the impact of the accuracy of the surrogate model on latency of the approaches, we previously evaluated our approaches with a synthetic probability of updates (cf. Sec.~\ref{ssec:accuracy-of-surrogate-model}).  To this end, we recorded the median latency for the different approaches for multiple simulation runs with different update probability.  We assumed that updates are uniformly distributed.  All other parameters are as our default configuration introduced in Section~\ref{ssec:setup}.

\begin{figure}[htb]
  \centering
  \includegraphics[width=\linewidth]{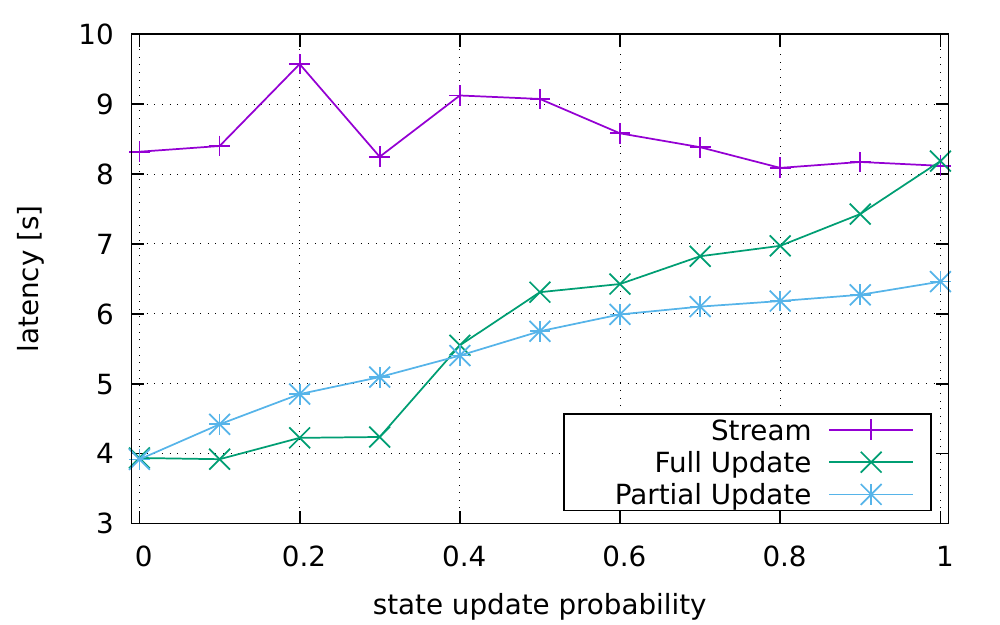}
  \caption{Latency over update probability.}
  \label{fig:updates}
\end{figure}

Figure~\ref{fig:updates} depicts latency over update probability of simulation states.  The latency of the stream approach is practically constant since it sends all states of the simulation over the network.  The full update approach is up to $2.1$ times faster than the stream approach for no updates but converges to the latency of the stream approach when all states require updating.  Performance of the partial update approach only changes gradually since more updates only marginally change the communication overhead.  For only few updates, the full update approach has the same performance than the partial update approach.  However, if all states require updating, the partial update approach is still $26\,\%$ faster compared to streaming and full update approach making the partial update approach the best option in a scenario with many small state updates.

\subsection{Impact of the Size of Partial Updates}

In addition to varying update probability, we also considered different sizes of partial updates.  To this end, we run the approaches and introduced fake updates.  Each state had a probability of $0.5$ to require an update.  Each update had a fixed size of violation points that require updating by partial updates.  All other parameters are as our default configuration in Section~\ref{ssec:setup}.

\begin{figure}[htb]
  \centering
  \includegraphics[width=\linewidth]{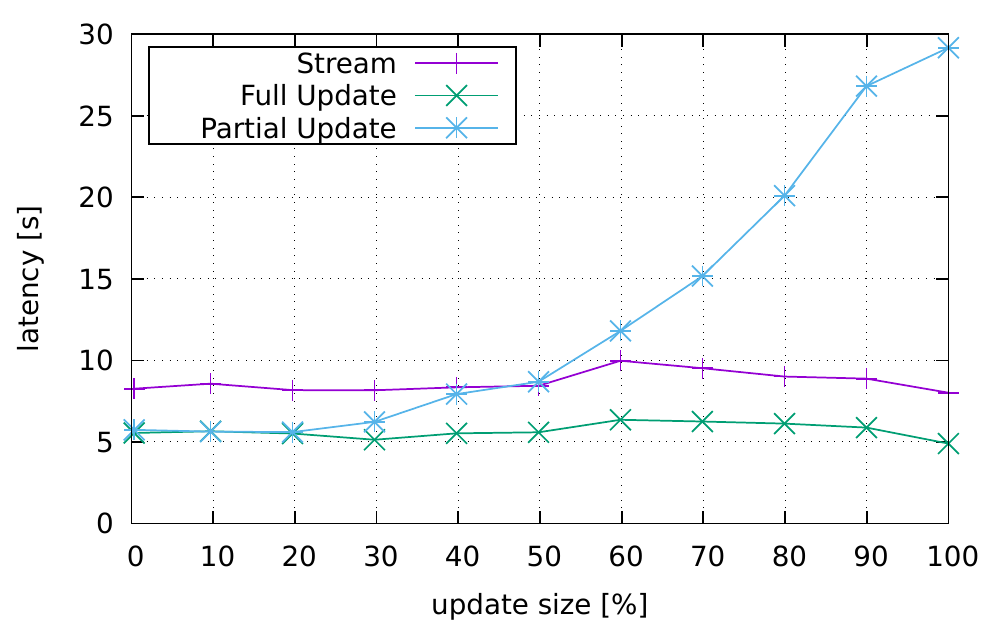}
  \caption{Latency over update size.}
  \label{fig:sizes}
\end{figure}

Figure~\ref{fig:sizes} depicts latency over update size.  As the size of the partial update does not effect the stream approach and full update approach, only the latency of the partial update approach gradually increases.  Latency is even higher than for the stream approach, since sending partial updates requires decoding of the position of the updated points.  This makes a partial update with all points updated bigger than a full update.  Additionally, the overhead for calculation of the ensemble Kalman filter is increased for more updates.  For more than $50$\,\% updates, streaming is more efficient than the partial update approach.  However, for up to $20$\,\%, the partial update approach provides results in the same time as the full update approach.  As shown in Section~\ref{ssec:accuracy-of-surrogate-model}, the percentage of violation points is typically below $20$\,\%.

\subsection{Impact of the Surrogate Problem Size}

Lastly, we want to measure the impact of different surrogate problem sizes.  If the surrogate problem grows, the stream approach has to communicate more data.  However, for the full update approach and the partial update approach, also the computational overhead is increased.  We want to measure the impact for different space discretization of the surrogate model.  All parameters are taken as described in Section~\ref{ssec:setup}.

\begin{figure}[htb]
  \centering
  \includegraphics[width=\linewidth]{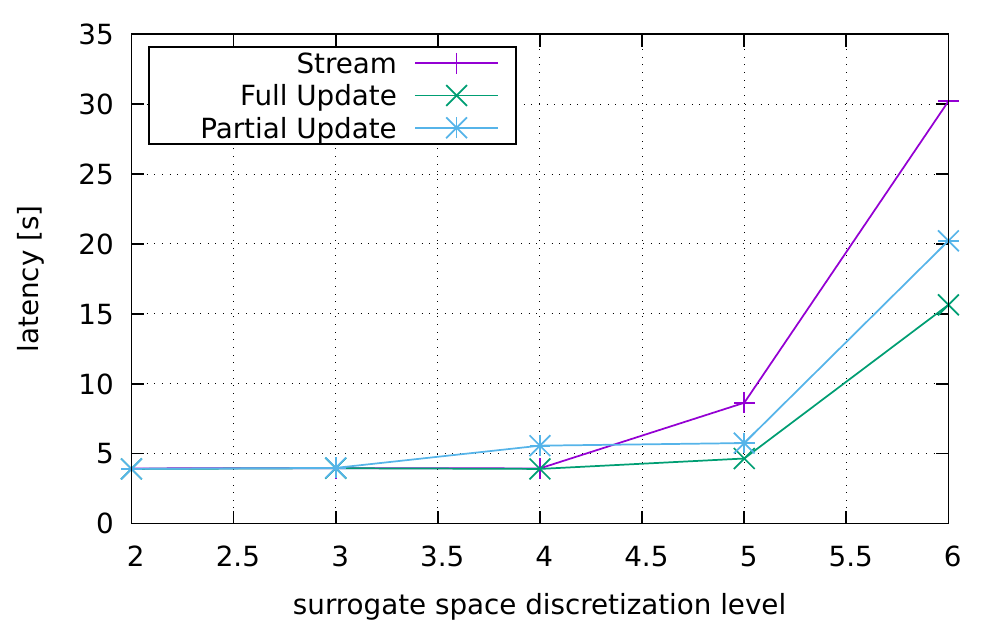}
  \caption{Latency over surrogate model space discretization level.}
  \label{fig:problem}
\end{figure}

Figure~\ref{fig:problem} shows the latency of the approaches over different discretization level.  Notice that for level 6, the reference model has the same space discretization as the surrogate model.  However, the surrogate model is implemented as implicit method, so the two models provide different results.  As for increased discretization level the number of unknowns grow exponentially, latency of the approaches grow linearly with increased number of unknowns.  However, the full update approach and partial update approach provide much better results for high discretization levels.  In particular, the full update approach provides a speedup of up to $1.9$, while the partial update approach is only $33$\,\% faster than the streaming approach.  Notice that the mobile device is limited to at most discretization level $6$ due to memory limitations.

\subsection{Conculsion}

Concluding the evaluations, full update approach and partial update approach are significantly better than streaming.  The full update approach is best for high data rates, high update size, low update probability, and high surrogate problem size.  The partial update approach is best in cases of low data rates, low update size, and high update probability.  Approaches can be combined by simply using the partial update approach for updates lower than $20$\,\% and otherwise send full updates to benefit from both update types.

\section{Conclusion \& Future Work}
\label{sec:conclusion-future-work}

In this article, we presented methods to provide results of resource-intensive simulations on resource-poor mobile devices.  The goal was to provide fast results with guaranteed quality.  To this end, our approaches compute the simulation in a user-defined reference quality on the server.  On the mobile device, a surrogate model providing lower quality at much fewer computation time will be executed.

We presented three approaches.  The first approach was to simply stream results to the mobile device in surrogate quality.  The second approach was to compute the surrogate model on the mobile device and the server.  The server detects if quality constraints are not fulfilled and send a full state update as correction to the mobile device.  In the third approach, we considered partial updates to reduce communication overhead.  To combine the surrogate state with partial updates, we use tools from data assimilation, namely the ensemble Kalman filter.  This is required to maintain mathematical properties of the simulation model.

All approaches were implemented and extensively evaluated on our test bed based on a Raspberry Pi and a connected server.  Evaluations showed that our approaches are able to provide fast simulation results, even in cases of low data rates.  Compared to our streaming approach, our approaches increase the performance of the system by up to over $13$ times.  The performance depends on the actual data rate and the size and frequency of required updates.

In the future, we plan to include uncertain sensor data in our approaches.  Additionally, we will research methods for signaling uncertainty to the user of the mobile device.

\subsection*{Acknowledgements}

The authors would like to thank the German Research Foundation (DFG) for financial support of the project within the Cluster of Excellence in Simulation Technology (EXC~310/2) at the University of Stuttgart.

\bibliographystyle{IEEEtran}
\bibliography{literature}

\begin{thebibliography}{10}
\providecommand{\url}[1]{#1}
\csname url@samestyle\endcsname
\providecommand{\newblock}{\relax}
\providecommand{\bibinfo}[2]{#2}
\providecommand{\BIBentrySTDinterwordspacing}{\spaceskip=0pt\relax}
\providecommand{\BIBentryALTinterwordstretchfactor}{4}
\providecommand{\BIBentryALTinterwordspacing}{\spaceskip=\fontdimen2\font plus
\BIBentryALTinterwordstretchfactor\fontdimen3\font minus
  \fontdimen4\font\relax}
\providecommand{\BIBforeignlanguage}[2]{{%
\expandafter\ifx\csname l@#1\endcsname\relax
\typeout{** WARNING: IEEEtran.bst: No hyphenation pattern has been}%
\typeout{** loaded for the language `#1'. Using the pattern for}%
\typeout{** the default language instead.}%
\else
\language=\csname l@#1\endcsname
\fi
#2}}
\providecommand{\BIBdecl}{\relax}
\BIBdecl

\bibitem{Dibak2015}
C.~Dibak, F.~D{\"u}rr, and K.~Rothermel, ``{Numerical Analysis of Complex
  Physical Systems on Networked Mobile Devices},'' in \emph{Proceedings of the
  12th IEEE International Conference on Mobile Ad hoc and Sensor Systems (MASS
  2015)}.\hskip 1em plus 0.5em minus 0.4em\relax IEEE, Oct 2015, pp. 280--288.

\bibitem{Dibak2017a}
C.~Dibak, A.~Schmidt, F.~D{\"u}rr, B.~Haasdonk, and K.~Rothermel,
  ``Server-assisted interactive mobile simulations for pervasive
  applications,'' in \emph{Proceedings of the 15th IEEE International
  Conference on Pervasive Computing and Communications (PerCom 2017)}.\hskip
  1em plus 0.5em minus 0.4em\relax IEEE, Mar 2017, pp. 111--120.

\bibitem{Cuervo2010}
E.~Cuervo, A.~Balasubramanian, D.-k. Cho, A.~Wolman, S.~Saroiu, R.~Chandra, and
  P.~Bahl, ``Maui: making smartphones last longer with code offload,'' in
  \emph{Proceedings of the 8th international conference on Mobile systems,
  applications, and services}.\hskip 1em plus 0.5em minus 0.4em\relax ACM,
  2010, pp. 49--62.

\bibitem{Ra2011}
M.-R. Ra, A.~Sheth, L.~Mummert, P.~Pillai, D.~Wetherall, and R.~Govindan,
  ``Odessa: enabling interactive perception applications on mobile devices,''
  in \emph{Proceedings of the 9th international conference on Mobile systems,
  applications, and services}.\hskip 1em plus 0.5em minus 0.4em\relax ACM,
  2011, pp. 43--56.

\bibitem{Chun2011}
B.-G. Chun, S.~Ihm, P.~Maniatis, M.~Naik, and A.~Patti, ``Clonecloud: Elastic
  execution between mobile device and cloud,'' in \emph{Proceedings of the
  Sixth Conference on Computer Systems}, ser. EuroSys '11.\hskip 1em plus 0.5em
  minus 0.4em\relax ACM, 2011, pp. 301--314.

\bibitem{Gordon2012}
M.~S. Gordon, D.~A. Jamshidi, S.~Mahlke, Z.~M. Mao, and X.~Chen, ``{COMET}:
  Code offload by migrating execution transparently,'' in \emph{Presented as
  part of the 10th {USENIX} Symposium on Operating Systems Design and
  Implementation ({OSDI} 12)}.\hskip 1em plus 0.5em minus 0.4em\relax {USENIX},
  2012, pp. 93--106.

\bibitem{Dibak2017b}
C.~Dibak, F.~D{\"u}rr, and K.~Rothermel, ``Demo: Server-assisted interactive
  mobile simulations for pervasive applications,'' in \emph{2017 IEEE
  International Conference on Pervasive Computing and Communications Workshops
  (PerCom Workshops)}.\hskip 1em plus 0.5em minus 0.4em\relax IEEE, Mar 2017,
  pp. 68--70.

\bibitem{Dibak2018a}
C.~Dibak, B.~Haasdonk, A.~Schmidt, F.~D{\"u}rr, and K.~Rothermel, ``Enabling
  interactive mobile simulations through distributed reduced models,''
  \emph{Pervasive and Mobile Computing}, vol.~45, pp. 19--34, 2018.

\bibitem{Berg2015}
F.~Berg, F.~D\"urr, and K.~Rothermel, ``Increasing the efficiency of code
  offloading through remote-side caching,'' in \emph{11th International
  Conference on Wireless and Mobile Computing, Networking and Communications
  (WiMob)}.\hskip 1em plus 0.5em minus 0.4em\relax IEEE, Oct 2015, pp.
  573--580.

\bibitem{Pandey2016}
P.~Pandey and D.~Pompili, ``Mobidic: Exploiting the untapped potential of
  mobile distributed computing via approximation,'' in \emph{International
  Conference on Pervasive Computing and Communications (PerCom)}.\hskip 1em
  plus 0.5em minus 0.4em\relax IEEE, Mar 2016, pp. 1--9.

\bibitem{Pandey2017}
P.~{Pandey} and D.~{Pompili}, ``Exploiting the untapped potential of mobile
  distributed computing via approximation,'' \emph{Pervasive and Mobile
  Computing}, vol.~38, pp. 381--395, 2017.

\bibitem{Law2015}
K.~Law, A.~Stuart, and K.~Zygalakis, \emph{Data assimilation}, ser. Texts in
  Applied Mathematics.\hskip 1em plus 0.5em minus 0.4em\relax Springer, 2015,
  vol.~62.

\bibitem{Evensen2003}
G.~Evensen, ``The ensemble kalman filter: theoretical formulation and practical
  implementation,'' \emph{Ocean Dynamics}, vol.~53, no.~4, pp. 343--367, Nov
  2003.

\bibitem{Houtekamer1998}
P.~L. Houtekamer and H.~L. Mitchell, ``Data assimilation using an ensemble
  kalman filter technique,'' \emph{Monthly Weather Review}, vol. 126, no.~3,
  pp. 796--811, 1998.

\bibitem{Keppenne2000}
C.~L. Keppenne, ``Data assimilation into a primitive-equation model with a
  parallel ensemble kalman filter,'' \emph{Monthly Weather Review}, vol. 128,
  no.~6, pp. 1971--1981, 2000.

\bibitem{Matsumoto1998}
M.~Matsumoto and T.~Nishimura, ``Mersenne twister: A 623-dimensionally
  equidistributed uniform pseudo-random number generator,'' \emph{ACM
  Transactions on Modeling and Computer Simulation}, vol.~8, no.~1, pp. 3--30,
  Jan. 1998.

\bibitem{Sumaray2012}
A.~Sumaray and S.~K. Makki, ``A comparison of data serialization formats for
  optimal efficiency on a mobile platform,'' in \emph{Proceedings of the 6th
  International Conference on Ubiquitous Information Management and
  Communication (ICUIMC)}.\hskip 1em plus 0.5em minus 0.4em\relax ACM, 2012,
  pp. 48:1--48:6.

\end{thebibliography}

\appendices

\section{Details About the Ensemble Kalman Filter}
\label{sec:details-about-kalman-filter}

In this appendix, we shortly describe the steps required to implement the ensemble Kalman filter (EnKF) for our partial update approach based on \cite{Houtekamer1998, Evensen2003}.  We first describe the input into the filter, the general idea, and then provide more details on the calculation.  Notice that while the EnKF can also be implemented for implicit methods, our implementation uses an explicit method as surrogate model.

The filter is based on a simulation model described by a matrix $M\in\mathbb{R}^{n\times n}$ and a problem specific procedure for perturbation of states.  After the perturbation, the mean of ensemble members should track the real state of the system, i.e., the reference simulation state.

For every system state, partial updates are provided by means of the updated values in a vector $u_{i+1}$ and a selection matrix $H_{i+1}$ that contains the position of the updates points.  Notice that $H_{i+1}$ and $u_{i+1}$ can be easily constructed from updates as list of tuples representation with $\{ (\textit{position}, \textit{value}), \dots \}$.

The idea of the EnKF is to track states $S_i$ and the uncertainty of the states as sample covariance of the ensemble members $e_i^{(j)}$.  Using the ensemble members, the so called Kalman gain $K$ is calculated for every state update.  The Kalman gain defines the influence of the update on the current state of the surrogate result and is calculated as $K = CH^T (HCH^T)^{-1}$, where $H^T$ represents the transpose of $H$.  The sample covariance $C$ is calculated as $C = E[(E - X)(E-X)^T]$, where $E[\cdot]$ denotes the expected value.  Variable $X$ should be a good guess on the reference simulation state, which is provided by the ensemble mean value.

\end{document}